# Low-loss and broadband non-volatile phase-change directional coupler switches

Peipeng Xu,*,†,‡,§ Jiajiu Zheng,†,‡ Jonathan Doylend,∥ and Arka Majumdar*,‡,⊥

‡Department of Electrical and Computer Engineering, University of Washington, Seattle, WA 98195, USA

§Laboratory of Infrared Materials and Devices, Advanced Technology Research Institute, Ningbo University, Ningbo 315211, China

∥Silicon Photonic Products Division, Intel Corporation, Santa Clara, CA 95054, USA

⊥Department of Physics, University of Washington, Seattle, WA 98195, USA

†These authors contributed equally to this work.

*To whom correspondence should be addressed. Email address: xupeipeng@nbu.edu.cn, arka@uw.edu.

## ABSTRACT

An optical equivalent of the field-programmable gate array (FPGA) is of great interest to large-scale photonic integrated circuits. Previous programmable photonic devices relying on the weak, volatile thermo-optic or electro-optic effect usually suffer from a large footprint and high energy consumption. Phase change materials (PCMs) offer a promising solution due to the large non-volatile change in the refractive index upon phase transition. However, the large optical loss in PCMs poses a serious problem. Here, by exploiting an asymmetric directional coupler design, we demonstrate non-volatile PCM-clad silicon photonic 1 × 2 and 2 × 2 switches with a low insertion loss of ~1 dB and a compact coupling length of ~30 μm while maintaining a small crosstalk less than −10 dB over a bandwidth of 30 nm. The reported optical switches will function as the building blocks of the meshes in the optical FPGAs for applications such as optical interconnects, neuromorphic computing, quantum computing, and microwave photonics.

Tremendous progress has been made in photonic integrated circuits (PICs) over the last two decades, revealing their potential to create photonic systems with small footprints, low power consumption, high-speed operation, and low-cost packaging. With PICs going fabless,[1] large-scale PICs have recently been reported, enabling systems with complexities far beyond classical benchtop optics.[2-4] Many of these PICs rely on programmable and generic photonic circuits[5-8] analogous to the field-programmable gate arrays (FPGAs) in electronics. Contrary to the scheme of application-specific PICs, where specific circuit architectures are designed to implement particular functions, such programmable PICs bring about far greater flexibility and effective cost reduction and thus will be a promising approach to realize applications such as routing fabrics in optical communication networks, reconfigurable logic gates in optical information processing, and multifunctional lab on a chip in optical sensing.[6]

Programmable optical cores employing a grid of Mach–Zehnder (MZ) switches have been demonstrated by several groups.[5,7,8] In these works, the on-chip optical switches can be reconfigured to the cross or bar state, forming one of the most fundamental and critical components in programmable PICs. Current optical switches in PICs, however, primarily rely on the weak modulation of the refractive index (usually $\Delta n < 0.01$) from the free-carrier dispersion[9,10] or thermo-optic[11] effects, resulting in a large footprint (several hundred micrometers) and high power consumption (typically several milliwatts). Resonator-based switches can help improve the modulation strength[12] but suffer from intrinsic narrow optical bandwidth as well as high sensitivity to fabrication imperfections and temperature fluctuations.[13] Moreover, as the switching mechanism is volatile, a constant power supply is required to maintain the switched state.

To overcome these fundamental limitations, phase-change materials (PCMs) have been proposed to provide strong modulation and nonvolatility for on-chip tunable optical devices[14,15] due to several unique properties: phase transition between amorphous and crystalline states with considerable modulation in electrical resistivity and optical constants ($\Delta n > 1$) over a broad spectral region,[16] state retention for years in no need of extra power,[17] fast and reversible switching of the states with nanosecond optical or electrical pulses,[18,19] high endurance up to $10^{15}$ switching cycles,[20] and excellent scalability.[21] Therefore, PCMs have emerged for a plethora of PIC applications such as optical switches,[15,22-28] optical modulators,[25,29] photonic memories,[30,31] and optical computing.[32,33]

Practical applications of programmable PICs require optical switches to have a multiport and broadband characteristic. Current experimental demonstrations of PCM-integrated switches, however, are either single-port[22,24,27] or narrow-band.[15,23,26] PCM-integrated MZ switches can afford broadband operation.[11] Unfortunately, their performance including the crosstalk (CT, defined as the contrast ratio between the two output ports) and insertion loss (IL) is dramatically sacrificed due to the large absorptive loss from crystalline PCMs.

Here, we demonstrate compact (~30 μm), low-loss (~1dB), and broadband (over 30 nm with CT < −10 dB) 1 × 2 and 2 × 2 switches using the PCM, $Ge_2Sb_2Te_5$ (GST), based on the previously built non-volatile programmable GST-on-silicon platform[15] and the asymmetric directional coupler (DC) switch design[28,34] bypassing the high loss associated with the crystalline state.

**RESULTS AND DISCUSSION**

Figure 1(a) shows the schematic of the 1 × 2 DC switch. The asymmetric coupling region consists of a normal silicon strip waveguide (SW) and a GST-on-silicon hybrid waveguide (HW) where a thin layer of GST is placed on silicon. When the GST is in the low-loss amorphous state, the optimized structure of the silicon SW and the HW can meet the phase-matching condition for TE polarization, leading to the cross state of the switch with a low IL [Figure 1(b)]. Once the GST is transformed to the lossy crystalline state, the phase-matching condition is significantly altered due to the strong modification of the mode in the HW induced by the dramatic difference of complex refractive indices between amorphous GST (aGST) and crystalline GST (cGST) (see Supporting Information). As a result, light is diverted away from the HW forming the bar state of the switch with low attenuation ensured by minimal optical field interaction with the lossy GST layer [Figure 1(c)].

To determine the widths of the waveguides ($w_s$, $w_c$) appropriately, we analyze the effective indices of the eigenmodes supported in the silicon SW and the GST-on-silicon HW (see Supporting Information). The simulations are performed using the frequency-domain finite-element method (COMSOL Multiphysics). The width of the silicon SW ($w_s$) is chosen as 450 nm to ensure single-mode operation. The width of the GST ($w_p$) is set to be 100 nm smaller than the core width of the HW ($w_c$), which can be easily achieved within the alignment precision of the electron-beam lithography (EBL). $w_c$ is optimally chosen as 420 nm so that the phase-matching condition could be satisfied for aGST. Therefore, the input TE-polarized light will be evanescently coupled to the cross port completely with an appropriate coupling length. Considering the trade-off between the coupling length ($L_c$) and the insertion loss in the crystalline state ($IL_{cGST}$) (see Supporting Information), the gap ($g$) between the two waveguides is chosen to be 150 nm while ensuring reliable fabrication of the coupling region. The coupling length given by $L_c = \lambda_0/2(n_{aGST1} - n_{aGST2})$ is thus calculated as compact as ~24 μm, where $n_{aGST1}$ and $n_{aGST2}$ are respectively the effective indices of the first order (even) and second order (odd) supermodes in the two-waveguide system, $\lambda_0 = 1550$ nm is the wavelength.

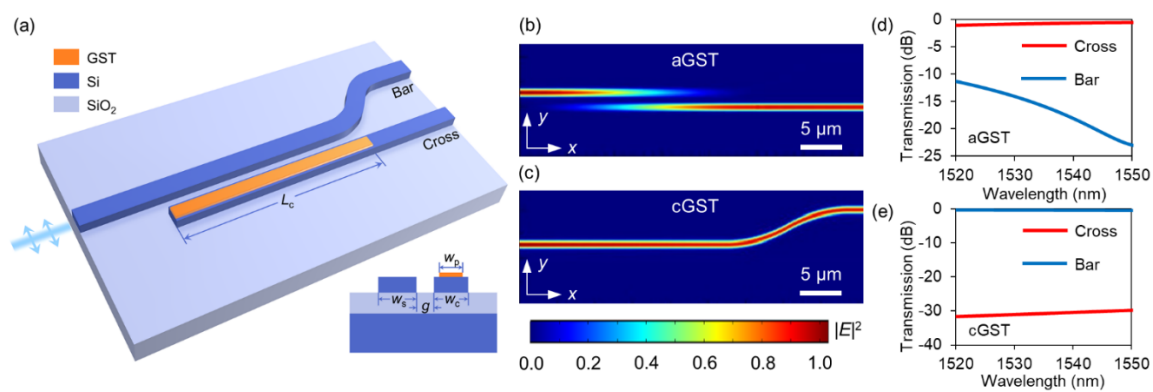

**Figure 1.** Design of the 1 × 2 DC switch. (a) Schematic of the switch. (b)-(c) Normalized optical field intensity distribution of the switch for (b) aGST and (c) cGST simulated by the 3D finite-difference time-domain method (Lumerical) at 1550 nm. (d)-(e) Calculated transmission spectra at the cross and bar ports for (d) aGST and (e) cGST.

Figure 1(d) and 1(e) show the calculated transmission spectral response of the 1 × 2 DC switch in both states. When the GST is in the amorphous state, the optical switch attains a small IL <1 dB and CT from −11 dB to −23 dB over the wavelength range of 1520~1550 nm. For the crystalline state, since almost no evanescent coupling occurs due to the phase mismatch, the spectral response to the input light is quite flat and broadband. The corresponding IL and CT are < 0.6 dB and < −29 dB across the whole wavelength range.

The devices were fabricated (see Supporting Information) using an SOI wafer with a 220-nm-thick silicon layer on top of a 3-μm-thick buried oxide layer. The pattern was defined via EBL and transferred to the top silicon layer by inductively coupled plasma etching. Deposition of 20-nm GST and 11-nm indium tin oxide (ITO to avoid GST oxidation) on the HWs was completed using a second EBL step followed by the sputtering and lift-off process. Figure 2(a) and 2(b) show the optical microscope and scanning electron microscope (SEM) images of the fabricated 1 × 2 DC switch. A false-colored SEM image of the coupling region is shown in Figure 2(c) where the GST layer is clearly resolved.

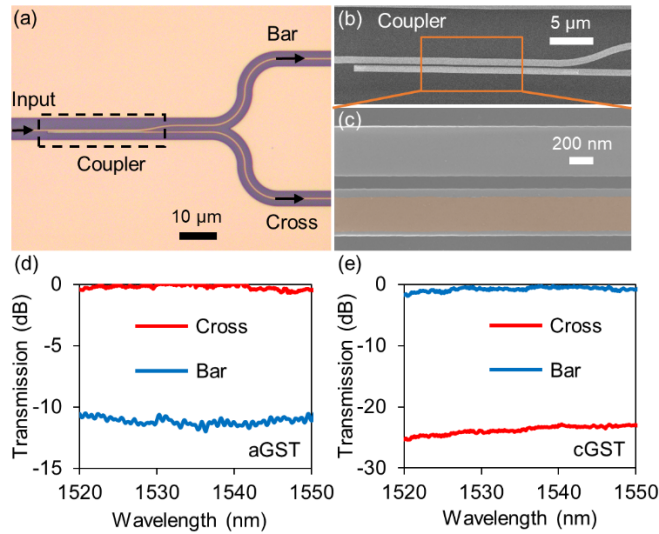

**Figure 2.** Experimental results of the 1 × 2 DC switch. (a) Optical microscope image of the fabricated switch. (b) SEM image of the switch. (c) An enlarged view of the coupling region highlighted by the orange rectangle in (b) with the GST false-colored. (d)-(e) Measured transmission at the cross and bar ports with the GST in the (d) amorphous and (e) crystalline states.

An off-chip optical fiber setup was used to measure the spectral response of the fabricated devices (see Supporting Information). For each device, we measured the transmission right after the deposition of the GST, which is initially in the amorphous state because of the low sputtering temperature. After that, rapid thermal annealing (RTA) of the chip at 200 °C for 10 mins in a $N_2$ atmosphere was performed to actuate the phase transition from aGST to cGST.[15] The measurement results of the 1 × 2 DC switch are shown in Figs. 2(d) and 2(e). For the wavelength range of 1520~1550 nm, the ILs were measured to be approximately 1 dB for both states and the CT was measured to be < −10 dB for aGST and < −22 dB for cGST. The discrepancy between the measured CTs and the design targets is primarily due to the fabrication-induced gap change and positional deviation of the GST layer owing to the limited alignment precision.

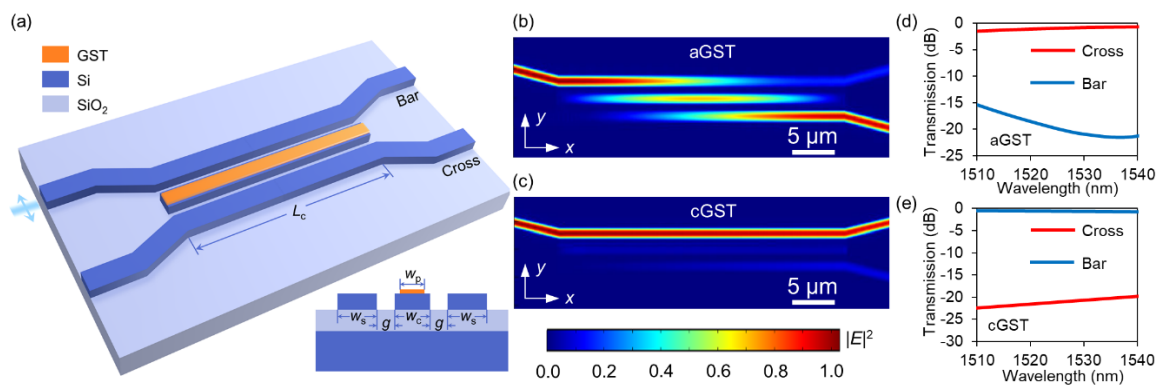

**Figure 3.** Design of the 2 × 2 DC switch. (a) Schematic of the switch. (b)-(c) Normalized optical field intensity distribution in the device for (b) aGST and (c) cGST simulated by the 3D eigenmode expansion method (Lumerical) at 1550 nm. (d-e) Calculated transmission spectra at the cross and bar ports for (d) aGST and (e) cGST.

We extend the DC design scheme to build a 2 × 2 switch. Figure 3(a) shows the schematic diagram of the 2 × 2 DC switch based on the three-waveguide DC. The operating principle of the proposed switch relies on the considerable mode modification of the TE-polarized supermodes in the three-waveguide system due to the GST phase transition. When the GST is in the amorphous state, the device functions as a three-waveguide DC and the complete power transfer could be achieved when the phase-matching condition (i.e. the effective indices of the three supermodes are evenly spaced) is satisfied. Thus, the input light couples to the low-loss GST-on-silicon HW and passes through the cross port [Figure 3(b)]. We study the light coupling mechanism by analyzing the supermodes in the coupling region (see Supporting Information). In this calculation, we adopt the same parameters used in the 1 × 2 switch with $w_s$ = 450 nm, $g$ = 150 nm, and $w_p = w_c$ − 100 nm. To meet the phase-matching condition, the width of the HW ($w_c$) is optimally chosen as 422 nm. Once the GST is crystallized, the three-waveguide system effectively boils down to two separated SWs because of the much higher effective index of the HW. In this case, only the even and odd supermodes can be supported in the coupler. The gap between the two SWs is $w_p + 2g$, resulting in a much larger coupling length ($L_{cGST}$). More specifically, when $g = 150$ nm, $L_{cGST}$ is calculated to be 516 μm while the coupling length for aGST ($L_{aGST}$) is 35 μm (see Supporting Information), leading to a large ratio of $L_{cGST}/L_{aGST} = 14.7$. Hence, after a specific coupling length ($L_c$, designed for the maximum transmission in the amorphous state, i.e. $L_{aGST}$), the input light is almost not cross-coupled but propagates directly to the bar port as if the central HW doesn't exist [Figure 3(c)]. This behavior can be further verified by the fact that the effective indices of the supermodes almost remain unchanged when $w_c$ changes as there is no field distribution in the GST-on-silicon HW (see Supporting Information).

Figure 3(d) and 3(e) show the simulated spectral response of the designed 2 × 2 DC switch for aGST and cGST when launching the light from one of the input ports. The device exhibits low ILs of <1 dB in both states when the wavelength varies from 1510 nm to 1540 nm. The bandwidth for achieving a CT less than −15 dB in the amorphous state and less than −20 dB in the crystalline state is more than 25 nm, enabling broadband switching operation.

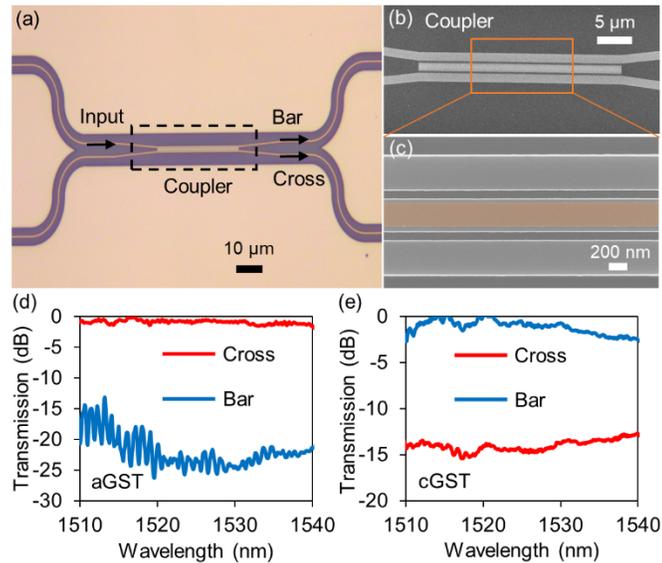

**Figure 4.** Experimental results of the 2 × 2 DC switch. (a) Optical microscope image of the fabricated switch. (b) SEM image of the switch. (c) An enlarged view of the coupling region

highlighted by the orange rectangle in (b) with the GST false-colored. (d)-(e) Measured transmission at the cross and bar ports with the GST in the (d) amorphous and (e) crystalline states.

Figure 4(a)-4(b) show the microscope and SEM images of the fabricated 2 × 2 DC switch. A false-colored SEM image of the coupling region is shown in Figure 4(c) where the GST layer is clearly resolved. The measured transmission spectra of the 2 × 2 DC switch are shown in Figure 4(d) and 4(e). For aGST, the IL was measured to be approximately 1 dB and the CT is less than ~−15 dB at the wavelength ranging from 1510 nm to 1540 nm, agreeing well with the simulation results. For cGST, the IL through the switch is approximately 1~2 dB and the CT is less than −10 dB with a bandwidth of over 30 nm. The degraded IL and CT compared with the simulation results are mainly attributed to the gap discrepancy due to the fabrication imperfection and positional deviation of the GST layer owing to the limited alignment precision.

**CONCLUSIONS**

By exploiting the high optical contrast of PCMs and asymmetric DC design, we demonstrated compact (~30 μm) non-volatile 1 × 2 and 2 × 2 switches with low-loss (~1 dB) and broadband (over 30 nm with CT < −10 dB) operations on the silicon photonic platform. With emerging wide-bandgap PCMs[35,36] and better fabrication, further improvement of the performance including IL and CT can be expected. From the volume of the GST needed in the switches, the reconfiguration energy for phase transition is estimated to be ~2 nJ,[15] only an order of magnitude larger than the thermodynamic limit.[20] Note that, due to the non-volatility of the GST, no more energy is required after switching. The availability of such on-chip non-volatile switching technology paves the way for optical FPGAs and sheds light on their applications including optical interconnects, neuromorphic computing, quantum computing, and microwave photonics.

**METHODS**

**Fabrication and optical characterization setup.** The designed on-chip optical switches were fabricated using the SOI wafer with a 220-nm-thick silicon layer on top of a 3-μm-thick buried oxide layer. The pattern was defined by a JEOL JBX-6300FS 100kV EBL system using a positive tone ZEP-520A resist and transferred to the silicon layer by inductively coupled plasma (ICP) etcher utilizing a gas mixture of $SF_6$ and $C_4F_8$. Next, a positive electron beam resist, PMMA was spun on the sample and a second EBL exposure was used to define the window for the GST deposition on the HWs. Finally, 20-nm GST and 11-nm indium tin oxide (ITO) were deposited using a magnetron sputtering system followed by a lift-off process. The on-chip devices were characterized by an off-chip optical fiber setup. The focusing sub-wavelength grating couplers[37] were fabricated at the input ports and output ports for fiber-chip coupling and polarization selectivity. The polarization of the input light was controlled to match the fundamental quasi-TE mode of the waveguide by a manual fiber polarization controller (Thorlabs FPC526). The straight single-mode waveguides with the same grating couplers were also fabricated on the same chip to normalize the spectra. A tunable continuous wave laser (Santec TSL-510) and a low-noise power meter (Keysight 81634B) were used to measure the performance of the fabricated devices.

**Acknowledgements**

The research was funded by the SRC grant 2017-IN-2743 (Fund was provided by Intel), NSF-EFRI-1640986, AFOSR grant FA9550-17-C-0017 (Program Manager Dr. Gernot Pomrenke), and UW Royalty Research Fund. P. X. is supported by National Natural Science Foundation of China (NSFC) (61505092, 61875099) and the Natural Science Foundation of Zhejiang Province, China (LY18F050005). Part of this work was conducted at the Washington Nanofabrication Facility / Molecular Analysis Facility, a National Nanotechnology Coordinated Infrastructure (NNCI) site at the University of Washington, which is supported in part by funds from the National Science Foundation (awards NNCI-1542101, 1337840 and 0335765), the National Institutes of Health, the Molecular Engineering & Sciences Institute, the Clean Energy Institute, the Washington Research Foundation, the M. J. Murdock Charitable Trust, Altatech, ClassOne Technology, GCE Market, Google and SPTS.

**Competing financial interests:**
The authors declare no potential conflicts of interest.

**Supporting Information**

The design of the 1 × 2 and 2 × 2 switches. This material is available free of charge *via* the Internet at http://pubs.acs.org.

# Supporting Information:

# Low-loss and broadband non-volatile phase-change directional coupler switches

Peipeng Xu,*,[†,‡,§] Jiajiu Zheng,[†,‡] Jonathan Doylend,[∥] and Arka Majumdar*,[‡,⊥]

[‡]Department of Electrical and Computer Engineering, University of Washington, Seattle, WA 98195, USA

[§]Laboratory of Infrared Materials and Devices, Advanced Technology Research Institute, Ningbo University, Ningbo 315211, China

[∥]Silicon Photonic Products Division, Intel Corporation, Santa Clara, CA 95054, USA

[⊥]Department of Physics, University of Washington, Seattle, WA 98195, USA

[†]These authors contributed equally to this work.

*To whom correspondence should be addressed. Email address: xupeipeng@nbu.edu.cn, arka@uw.edu.

**S1. Design of 1 × 2 switch**

An asymmetrical directional coupler (DC) consisting of a normal silicon strip waveguide (SW) and a GST-on-silicon hybrid waveguide (HW) is utilized for the design of the 1 × 2 switch. The widths ($w_s$, $w_c$) of the silicon SW and the HW are optimally chosen to make their mode effective indices equal when the GST is in the amorphous state. Fig. S1(a) shows the calculated effective indices of the fundamental modes of the silicon SW and the HW with respect to their widths. The right side shows the electrical field distribution for the TE mode of the HW with GST in the amorphous (aGST) and crystalline (cGST) state. The width of the silicon SW ($w_s$) is chosen as 450 nm to ensure single-mode operation. The width of the GST ($w_p$) is set to be 100 nm smaller than the core width of the HW ($w_c$), which can be easily achieved within the alignment precision of the electron-beam lithography (EBL). As a result, the optimal $w_c$ is 420 nm to achieve strong coupling between two waveguides, which can be verified in Figs. S1(b) and S1(c). Therefore, the input TE-polarized light will be evanescently coupled to the cross port completely with an appropriate coupling length. On the contrary, with the GST switched to the lossy crystalline state, the huge effective index contrast between the two waveguides indicates a serious phase mismatch, resulting in two isolated modes as shown in Figs. S1(d) and S1(e).

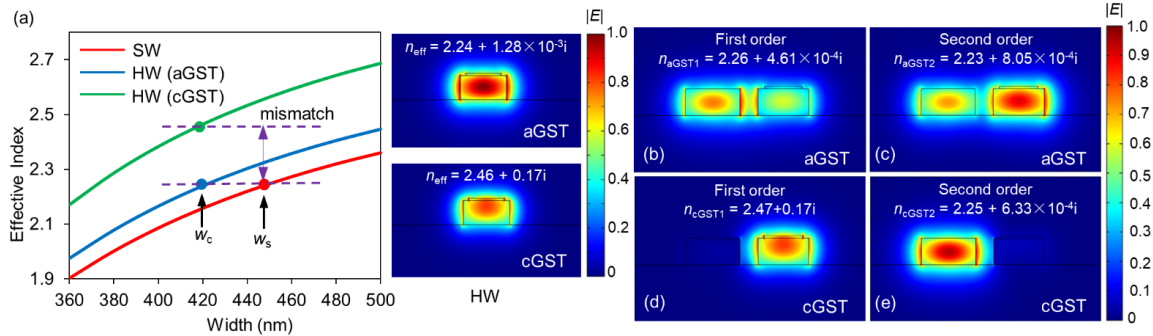

**Figure S1.** (a) Left: effective indices of the fundamental modes in the silicon SW and the HW. Right: the normalized electrical field profiles of the TE mode with GST in the amorphous and crystalline state. The widths of the SW and the HW are $w_s$ = 450 nm and $w_c$ = 420 nm, respectively. (b)-(e) Normalized electrical field profiles of the supermodes in the two-waveguide system when the GST is in (b-c) amorphous and (d-e) crystalline state. All the modes are simulated for TE polarization at 1550 nm.

Next, the characteristics of the coupling region are studied when varying the gap ($g$) between the silicon SW and HW. There exists a trade-off between the coupling length ($L_c$) and the insertion loss in the crystalline state ($IL_{cGST}$). $IL_{cGST}$ is introduced to access the loss performance for cGST and is given by

$$IL_{cGST} = L_c \times \alpha_{cGST2}, \qquad (S1)$$

where $L_c$ and $\alpha_{cGST2}$ denote the optimal coupling length corresponding to a specific $g$ and attenuation loss of the second order supermode [see Fig. S1(e)]. Figure S2 shows the $L_c$ and $IL_{cGST}$ as a function of the gap. As the gap increases, the coupling length in the amorphous state increases, while the insertion loss in the crystalline state decreases due to the weaker evanescent coupling between the two waveguides. Considering the trade-off and fabrication difficulty, a moderate $g$ of 150 nm is chosen.

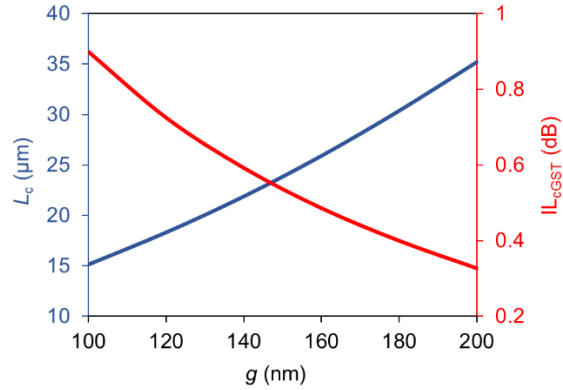

**Figure S2.** Coupling length for the maximum transmission in the amorphous state and insertion loss in the crystalline state as a function of $g$.

**S2. Design of 2 × 2 switch**

The three-waveguide DC consisting of two identical silicon strip waveguides separated by a GST-on-silicon hybrid waveguide is used to implement the 2 × 2 switch. Figure. S3(a)-S3(e) shows the transverse electric field distribution of all the supermodes supported by the three-waveguide coupler at 1550 nm in the amorphous and crystalline state. For aGST, the three-waveguide coupler supports two symmetric modes and one antisymmetric mode, as shown in Figs. S3(a)-S3(c). For cGST, only even and odd supermodes can be supported in the coupler, as shown in Figs. S3(d) and S3(e). In this state, the device can be simply regarded as a two-core DC. The dependence of the effective indices on wc is shown in Fig.

S3(f). $n_{aGST1}$, $n_{aGST2}$, and $n_{aGST3}$ represent the three effective indices of the supermodes in the amorphous state. To achieve the maximum power transfer efficiency to the cross port,[1,2] the effective indices of the supermodes need to meet the phase-matching condition given by $n_{aGST1} + n_{aGST2} = 2n_{aGST3}$. As $w_c$ increases, $n_{aGST1}$ and $n_{aGST2}$ increase and the phase-matching condition can be satisfied when choosing $w_c$ = 422 nm. Similarly, $n_{cGST1}$ and $n_{cGST2}$ represent the two effective indices of the supermodes in the crystalline state. They almost remain constant when $w_c$ changes as there exists no field distribution in the hybrid GST-on-silicon waveguide.

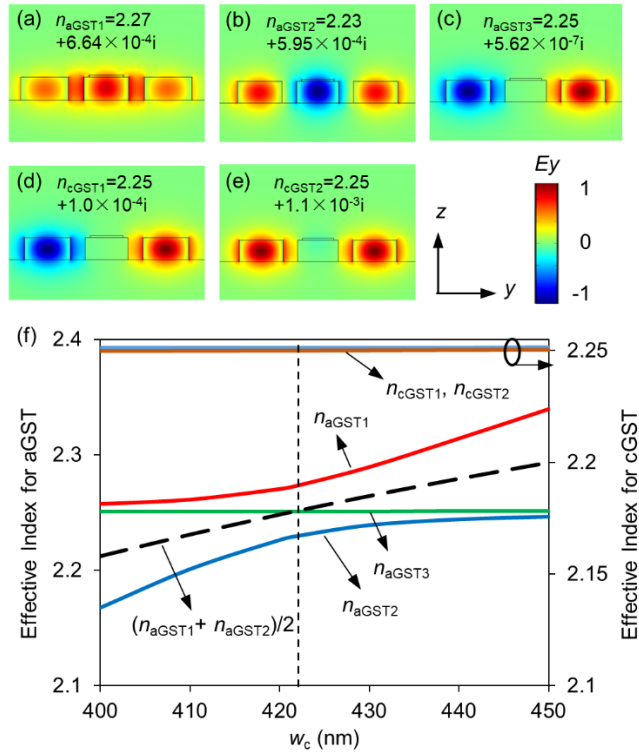

**Figure S3.** (a)-(e) Normalized $E_y$ field profiles of the supermodes in a three-waveguide system when GST is in the (a)-(c) amorphous and (d)-(e) crystalline state. (f) Effective indices of five supermodes guided in the three-waveguide coupler as a function of $w_c$ when $g$ = 150 nm, $w_s$ = 450 nm, $w_p = w_c - 100$ nm. The maximum coupling efficiency can be achieved when $w_c$ = 422 nm, i.e., $n_{aGST3} = (n_{aGST1} + n_{aGST2})/2$. All the modes are simulated for TE polarization at 1550 nm.

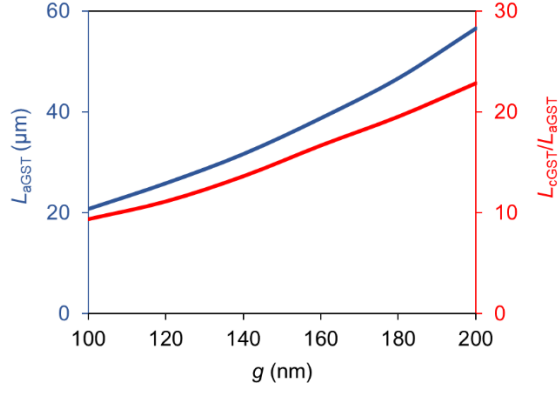

**Figure S4.** $L_{aGST}$ and $L_{cGST}/L_{aGST}$ as a function of $g$.

Next, the coupling characteristics are studied when varying the gap ($g$). Fig. S4 shows the corresponding coupling length for aGST ($L_{aGST}$) and the ratio of the two coupling lengths for cGST and aGST ($L_{cGST}/L_{aGST}$). As the gap increases, $L_{aGST}$ and $L_{cGST}/L_{aGST}$ increase simultaneously, implying a smaller crosstalk and insertion loss due to the cross-coupling but at the cost of a larger device length. Considering this performance trade-off and fabrication feasibility, we choose $g = 150$ nm. The coupling length of the three-waveguide coupler for both states can be calculated by

$$L_{aGST} = \frac{\lambda_0}{2(n_{aGST1} - n_{aGST3})} = \frac{\lambda_0}{2\Delta n_{aGST}}, \quad \text{(S2)}$$

$$L_{cGST} = \frac{\lambda_0}{2(n_{cGST1} - n_{cGST2})} = \frac{\lambda_0}{2\Delta n_{cGST}}. \quad \text{(S3)}$$

When $g = 150$ nm, $L_{aGST}$ and $L_{cGST}$ are calculated to be 35 μm and 516 μm, respectively, resulting in a large ratio of $L_{cGST}/L_{aGST} = 14.7$. Hence, it is expected that the switching operation of the two states can be achieved if we choose a suitable coupling length for the amorphous state.

We simulated the transmission spectra of the 2 × 2 switch using the eigenmode expansion method (Lumerical's MODE Solution). This method is especially suitable for our 2 × 2 switch due to the invariant structure in the propagation direction of light and will significantly reduce the calculation time by reusing the scattering matrices of similar structure sections.[3] The grid size of the mesh in our simulation is set as $\Delta x = \Delta y = \Delta z = 20$ nm. To ensure the accuracy of the field distribution, a grid size of 2 nm is used to override the GST domain.